# A Parsing Scheme for Finding the Design Pattern and Reducing the Development Cost of Reusable Object Oriented Software

K. M. Azharul Hasan, Mohammad Sabbir Hasan
Computer Science and Engineering Department
Khulna University Of Engineering and Technology(KUET)
Khulna 9203, Bangladesh.
azhasan@gmail.com, shabbir_cse03@yahoo.com

## ABSTRACT

*Because of the importance of object oriented methodologies, the research in developing new measure for object oriented system development is getting increased focus. The most of the metrics need to find the interactions between the objects and modules for developing necessary metric and an influential software measure that is attracting the software developers, designers and researchers. In this paper a new interactions are defined for object oriented system. Using these interactions, a parser is developed to analyze the existing architecture of the software. Within the design model, it is necessary for design classes to collaborate with one another. However, collaboration should be kept to an acceptable minimum i.e. better designing practice will introduce low coupling. If a design model is highly coupled, the system is difficult to implement, to test and to maintain overtime. In case of enhancing software, we need to introduce or remove module and in that case coupling is the most important factor to be considered because unnecessary coupling may make the system unstable and may cause reduction in the system's performance. So coupling is thought to be a desirable goal in software construction, leading to better values for external software qualities such as maintainability, reusability and so on. To test this hypothesis, a good measure of class coupling is needed. In this paper, based on the developed tool called Design Analyzer we propose a methodology to reuse an existing system with the objective of enhancing an existing Object oriented system keeping the coupling as low as possible.*

## 1. INTRODUCTION

There is increasing pressure on software developers to produce quality software in as short time as possible. This necessitates the reuse of previously developed or commercially available software elements to expedite the development process. The most common form of reuse is the reuse of code in a fine-grain manner such as objects in the object-oriented paradigm or a large-grain manner such as components in the component oriented paradigm [1]. A severe problem is encountered, however, is the quickly increasing complexity of such systems and the lack of adequate criteria and guidelines for good designs. To cope with this problem, it is imperative to better understand the properties and characteristics of object-oriented systems.

Reusability of software is considered as crucial technical precondition to improve the overall software quality and reduce production and maintenance cost [2]. Software components are supposed to be better reusable and more flexible compared to conventionally developed software. Unfortunately, the benefits associated technology has their price [3]. Poor documentation of the code makes studying and understanding details painful and time consuming for reusing that software. Abstracting design details from the source code help to understand the implementation details reuse it in efficient manner. Sometimes it becomes even impossible to understand particular document without the design details. Design patterns provide ways to structure software components into systems that are flexible, extensible, and have a high degree of reusability. Design patterns are an attempt to capture expertise in building object-oriented software that describes solution to a recurring design problem in a systematic and general way. Gamma at el [4] defines design patterns as "descriptions of communicating objects and classes that are customized to solve a general design problem in a particular context." A design pattern names, abstracts, and identifies the key aspects of a common design structure that makes it useful for creating a reusable object-oriented design. The design pattern identifies the participating classes and their instances, their roles and collaborations, and the distribution of responsibilities. Object-oriented design patterns typically show relationships and interactions between classes or objects, without specifying the final application classes or objects that are involved[3][5]. In this paper a methodology is proposed to find the design pattern of the software from the source code. This would provide several goals in software construction leading to better values for external attributes such as maintainability, reusability, and reliability. To find the design pattern it is important to know the interactions between the internal components of the software. Different interactions are pointed out and described to get the interactions.

Coupling characterizes a module's relationship to other modules of the system. It measures the interdependence of two modules[4]. Coupling measures the strength of physical relationships among the items that comprise an object. Strong coupling makes a system more complex, highly inter-related modules are harder to understand, change or correct. Designing the systems with the weakest possible coupling between modules can reduce complexity[7]. To validate the interactions we have selected three different types of industrial software for case study. For analyzing the coupling measures of object oriented systems based on different interactions of the classes, a parser has been developed named "*Design Analyzer*" to find the design pattern of the system. Such design analyzer is useful to get internal software architecture of software developed in Object Oriented Paradigm[1]. Finally, using principal component analysis [7] the measures are analyzed for selecting the most responsive coupling measure.

## 2. CASE STUDY

In this research work the analysis was performed on four industrial softwares developed in Object Oriented paradigm. A tool named "*Design Analyzer*" is developed as a part of this research work for parsing the source code to get the design pattern. The selected softwares were developed prior to the analysis and no modification was done during the analysis. For proceeding in an efficient manner the process of analyzing the source codes should follow the syntax of the programming language. Brief descriptions of these softwares are given here:

**Software 1**-Turtle Chat: This is chatting software like Yahoo! Messenger. This software can send and receive instant messages over Internet Protocol. There are two parts of this software: Server Part and Client Part. Here the Client side of contains 22 user defined classes and the Server side contains 4 user defined classes.

**Software 2**-Com Chat: This is also chatting software, which sends and receives data through the communication port of a personal computer. The software uses Java Communication API. The main purpose of this software is to simulate the seven layers of the OSI Model. The software implements Broadcasting, CheckSum and Routing techniques. This software contains 22 user defined classes.

**Software 3**-Admission Test Management System: This is mainly database software which can be used for admission test management of any university by storing information of the applicants, information of allowed candidate for exam, merit list of selected candidates, waiting list and so on. This software contains 13 user defined classes.

## 3. RELATED WORKS

In OO paradigm coupling describes the interdependency between methods and between object classes, respectively [6][7]. Stevens *et al.*, who first introduced coupling, in the context of structured development techniques, define coupling as "the measure of the strength of association established by a connection from one module to another" [9]. Braind et al. defined some properties to be satisfied by the coupling measure for empirical validation [10]. There are differences between necessary and unnecessary coupling. The rationale is that without any coupling the system is useless. Consequently, for any given software solution there is a basic or necessary coupling level. Such unnecessary coupling does indeed needlessly decrease the reusability of classes. There is also static and dynamic coupling measure for object oriented systems. As the polymorphic method invocation is determined by run time, the coupling on this method belongs to dynamic coupling [11]. It is not very much clear what the potential uses of existing measures are and how different coupling measures could be used in a complementary manner to obtain a more detailed picture of the coupling in an object-oriented system. Several authors have tried to address this problem by introducing frameworks to characterize different approaches to coupling and the relative strengths of it. There are some existing and quite different frameworks for object-oriented coupling. Eder et al. identify three different types of relationships[12]. These relationships, interaction relationships between methods, component relationships between classes, and inheritance between classes, are then used to derive different dimensions of coupling.

Hitz and Montazeri [13] characterize coupling by defining the state of an object (the value of its attributes at a given moment at run-time), and the state of an object's implementation (class interface and body at a given time in the development cycle). From these definitions, they derive two "levels" of coupling, Class level coupling (CLC), represents the coupling resulting from implementation dependencies between two classes in a system during the development lifecycle and Object level coupling (OLC), represents the coupling resulting from state dependencies between two objects during the run-time of a system.

An approach to measure coupling in object-based systems [14] such as those implemented in C++ by expanding it to include inheritance and friendship relations between classes. This framework concentrates on coupling as caused by interactions that occur between classes[15].

All the coupling measures mentioned in this paper are developed for using as metric suite. In this paper we defined the interactions to find the design pattern of software. Designing software is phase which is normally done before the implementation of software. But in this paper the design pattern of a software has been recovered analyzing the existing source code. This design pattern will be used to reuse the existing source code to modify or extend the software. This is a basic difference between other software metric related works. Beside, In this a methodology is proposed to add a new module in a existing software.

## 4. A METHODOLOGY TO FIND THE DESIGN PATTERN FOR JAVA BASED OBJECT ORIENTED SYSTEM

Before we go any further, it is imperative to first discuss the concept of a design pattern. Gamma at el defines design patterns as "descriptions of communicating objects and classes that are customized to solve a general design problem in a particular context." A design pattern names, abstracts, and identifies the key aspects of a common design structure that makes it useful for creating a reusable object-oriented design. The design pattern identifies the participating classes and their instances, their roles and collaborations, and the distribution of responsibilities. Object-oriented design patterns typically show relationships and interactions between classes or objects, without specifying the final application classes or objects that are involved. But beyond a description of the problem and its solution, software developers need deeper understanding to tailor the solution to their variant of the problem. Hence a design pattern also explains the applicability, trade-offs, and consequences of the solution. It gives the rationale behind the solution, not just a pat answer. In this paper the proposed *Design Analyzer* finds the interactions, roles, collaborations and relationships of the software in a graphical format.

In the subsequent sub sections the interactions between objects is found out for Java based programs. In Section 3 these interactions are used to develop the algorithm for *Design Analyzer*. The term object is used to mean a module through out the paper.

### 4.1 Types of Inter-module Interactions That Occur in Java

There different ways that are commonly found for interactions between the classes in a Java based object-oriented software are described in this section. These inter module relationships are as follows:
1. Inter-module relationship through Return Type
2. Inter-module relationship through Argument Passing in a member function
3. Inter-module relationship through Object Declaration
4. Inter-module relationship through Inheritance

From these four types of relationship, we have defined two types of interactions:
   a) Operation-Operation (O-O) interaction
   b) Class-Class interaction (C-C) interaction

Operation-Operation (O-O) interaction: The first two categories (relationship through return type and argument passing) above are included in O-O interaction. Hence we defined O-O interaction as follows

**Definition 1:** (Operation-Operation Interaction): The Operation-Operation interaction (O-O) is defined as the interaction between two operations of two or more different objects or classes. Let $O_C$ be an operation of class C. There is an operation-operation interaction between classes C and D, if class D is the type of a parameter of operation $O_C$ or class D is the return type of $O_C$.

Class-Class interaction (C-C) interaction: The last two categories above (relationship through object declaration and inheritance) are included in C-C interaction. Hence we defined C-C interaction as follows

**Definition 2:** (Class-Class Interaction): The Class-Class interaction (C-C) is defined as the interaction between two classes if any one of the above two interaction occurs (i.e. interaction through object declaration or inheritance). Let C and D be two classes of an object oriented system. There is a C-C interaction between the classes C and D, if an object $O_d$ of D is declared inside class C or D is derived from class C through inheritance.

**Definition 3:** (Interaction Graph ) The Interaction Graph <G,E> of a software is a graph where each node G represents a class of the system and there is an edge E between two nodes $G_1$ and $G_2$ if there is an interaction (O-O, C-C) between the two classes. Our developed and proposed Design Analyzer is an Interaction Graph.

### 4.2 The Parsing Scheme to Find the Design Pattern

In this section the two types of interactions defined in Section 2 (C-C and O-O) is parsed in java based programs to develop the *Design Analyzer*. Although the examples are in Java but it can easily be extended in any object oriented language.

### 4.2.1  Parsing for the O-O Interaction

*Relation through Argument Passing*

Before proceeding we have to know the format of how classes can be inter-related through argument passing in functions. Let's consider two classes Class A and Class B. Let a is the object of Class A which is declared in the scope of Class B. In java this can happen in one of the following fashion:

(1) Class B{
    access-modifier static A function-name (argument list)
    }

(2) Class B{
    access-modifier  final A function-name (argument list)
    }
(3) Class B{
    access-modifier  A function-name (argument list)
    }

Here access-modifier sits for indicating public, private or protected and argument list represents variables of any data type. So from here we see that Class B is the container class because it contains a function that uses the object of another class as argument. During parsing the source codes of Class B if we find a statement like:

*access-modifier static A function-name (argument list) or  access-modifier final A function-name (argument list) or*
 *access-modifier A function-name (argument list);*

then we can come to the decision that Class B is related to Class A through the object a and c as argument of the function function-name.

*Relation through Return Type of function*

Considering  two classes Class A and Class B. Let a is the object of Class A which is declared in the scope of Class B. In java this can happen in one of the following fashion:

(1) Class B{
    access-modifier static return-type function-name (A a, A c);
    }

(2) Class B{
    access-modifier  final return-type function-name (A a, A c)
    }

(3)  Class B{

access-modifier  return-type function-name (A a, A c)
    }

Here access-modifier sits for indicating public, private or protected and return type specifies value of any data type or void. So from here we see that Class B is the container class because it contains a function that uses the object of another class as argument. During parsing the source code of Class B if we find a statement like access-modifier static return-type function-name (A a, A c) or  access-modifier final return-type function-name (A a, A c) or  access-modifier return-type function-name (A a, A c);

then we can come to the decision that Class B is related to Class A through the object a and c as argument of the function function-name.

### 4.2.2 Parsing for the C-C interaction

*Relation through object declaration*

Before proceeding we have to know the format of how classes can be inter-related through object declaration in Java. Let's consider two classes Class A and Class B. Let a is the object of Class A which is declared in the scope of Class B. In java this can happen in the following fashion.

Class B {

     A  a = new A( );
    }

So from here we see that Class B is the container class because it contains the object of another class. During parsing the source code of Class B if we find a statement like A a = new A ( ); then we can come to the decision that Class B is related to Class A through the declaration of the object a.

*Relation through Inheritance*

Let's consider two classes Class A and Class B. Let a is the object of Class A which is declared in the scope of Class B. In java this can happen in one of the following fashion:

(1) Class B extends A{
            // body of Class B
    }
(2) Class B implements A {
            // body of Class B
    }

So from here we see that Class B inherits Class A. During parsing the source code of Class B if we find a statement like Class B implements A or Class B extends A then we can come to the decision that Class B is related to Class A through inheritance.

## 5. AN ANALYSIS OF DESIGN PATTERN USING DESIGN ANALYZER

We have developed tool using java for analyzing the source code of an object oriented system to get the design pattern of the system. We named it *Design Analyzer*. Through out the paper where we used the word *Design Analyzer* we mean the developed software. The *Design Analyzer* implements the parsing scheme described in Section 4. The input of the system is the source code of an Object Oriented program and output is the graphical representation (See Fig. 1) of the design pattern of the software. The design pattern is a graph <G,E> where each node G represents a class of the system and there is an edge E between two nodes if there is an interaction (O-O, C-C) between the two classes. We have considered only the user defined classes that are found in the source code. This is because our next goal is to add a new module in the system so that we can reuse the existing system with minimum modification.

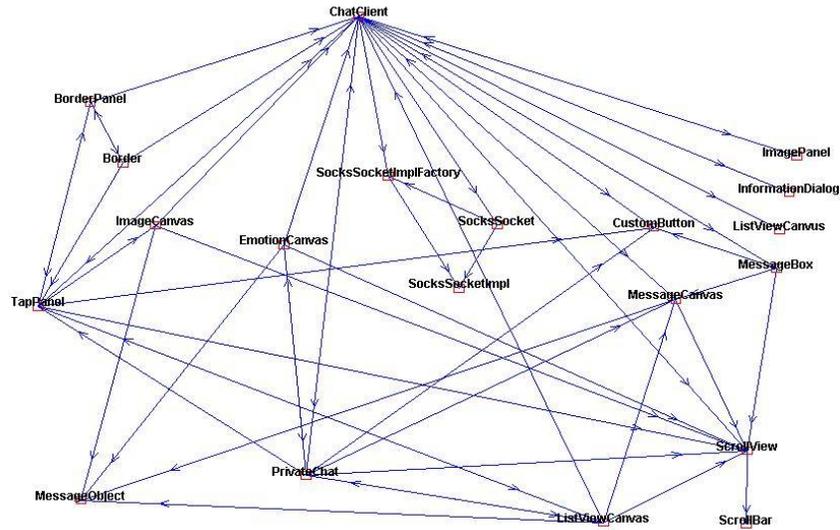

Fig 1: Graphical Representation of the relationship among the User Defined Classes of software 1

Fig.1 shows the design pattern found using the *Design Analyzer* for software 1 described in Section 1 from the figure we can see that there are 19 user defined classes in the system. Among them the class *ChatClient* is very much coupled with other classes. The *ChatClient* class has 16 coupling relations with others. We can see that there is no class isolated in the system. Hence this is a criterion of good design. But the distribution of coupling is based on only three classes namely *ChatClient, ScrollView* and *Tappanel*. This is a sign of low maintainability. Because if the class *ChatClient* fails for any reason most of the classes will be affected. Fig. 2 shows the design pattern of software 2. From Fig.3 we see that the average coupling distribution is similar to the classes. Fig. 3 shows the design pattern of software 3. From the figure we see that the software is poorly designed. Almost all the classes are coupled with one class namely *Admission*. If this class fails or has a bug then the whole software will work poorly. In conclusion we can say that to reuse a software it is very important to know about the design pattern of the software. If it is poorly designed then it might be error prone for reuse in the future.

## 6. The Coupling Measures

In this research work the following couplings metrics have been chosen for the analysis of inter module dependencies. A brief definition of the measures is given here. All the measures determine the coupling between components. A survey of the metrics can be found in [16][17].

**1. Number of used classes by dependency relation (NUCD)**: This measure is used to count the total number of distinct classes with whom a particular class is creating dependency relation. Only one evidence for dependency relation would be enough, caused by any of dependency types (e.g. parameter, local variable, return type) to recognize the dependency between two classes.

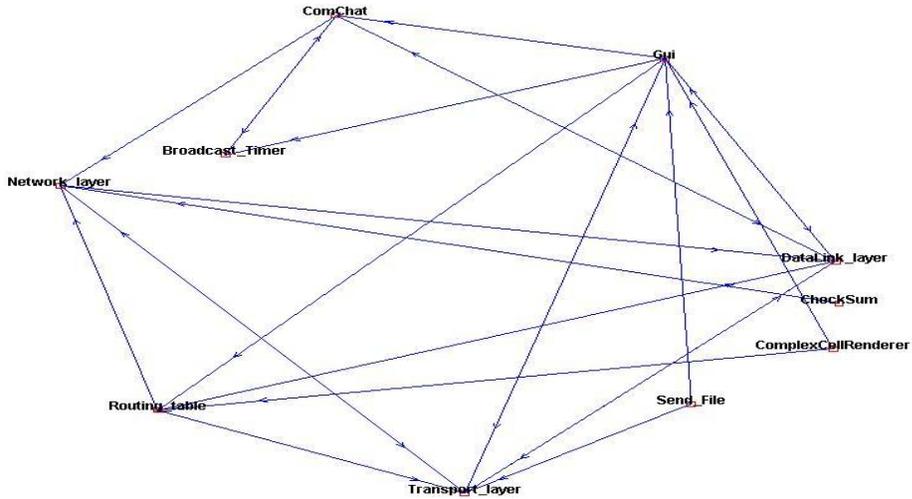

Fig 2: Graphical Representation of the relationship among the User Defined Classes of Software 2

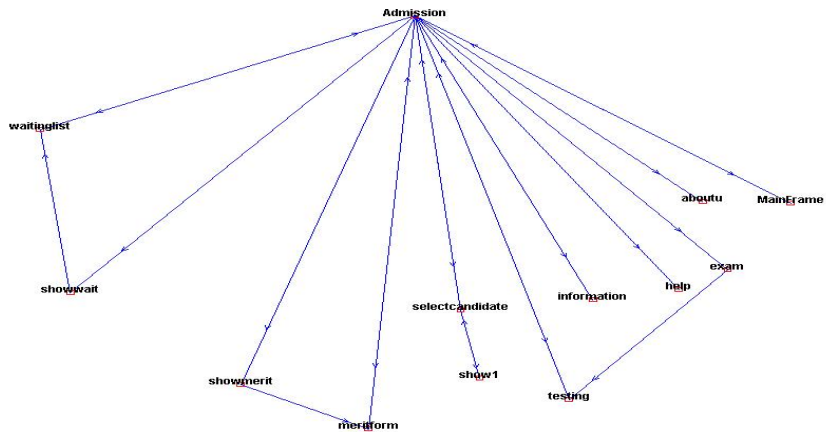

Fig 3: Graphical Representation of the relationship among the User Defined Classes of Software 3

**2. Total number of evidences for "Used classes by dependency relation" (TNUCD)**: This measure is used to count total number of evidences for a particular class of "Used classes by dependency relation." All types of dependencies (e.g. parameter, local variable, return type) will be used to count such evidences.

**3. Number of user classes for a class through dependency relation (NUCC):** This measurement represents the total number of distinct classes who are using a particular class through dependency relations.

**4. Total number of evidences for "User classes through dependency relation" (TNUCC):** This measurement counts the total number of usage evidences of a particular class by the other classes in OO design.

**5. Class Coupling:** The Class Coupling (CLC) is the summation of Client Coupling and Server Coupling of the class. It is the measure of the summation of out degree and in degree of a node in Class Class Interaction Graph(CCIG).

Class Coupling= Client Coupling (CC) + Server Coupling (SC).

The number of Coupling Relations for which a class is a client to other class is called Client Coupling for the class. It is the measure of out degree of a node in CCIG. The number of Coupling Relations for which a class is a server to other class(s) is called Server Coupling for the class. It is the measure of in degree of a node in CCIG.

**6. Visible Member:** The measure "Visible Member" shows the amount of members (attributes and methods) visible to other class numerically. This measure is used to find the over all members which can be called or used by other classes. Visible members are the most required criteria for direct coupling

## 7. Criteria of Measuring Coupling

We have developed an Interaction Graph (See Definition 3) and coupling occurs due to this interaction, hence among the coupling measures to select an effective one is necessary and useful. The methodology of Principal Component Analysis[7] has been adopted to select the most responsive coupling measure for a system. When we want to add a module we need to find a class that is less responsive. From graphical analysis we take decision to add a module in which the interaction graph has fewer edges and in this section we show some experimental result to prove the idea.

### 7.1 Experiment Design

To make a study of coupling measure we want to determine the best coupling metrics defined above. We are going to apply these measures on the software under observation. The above measures are applied on these softwares through principal component analysis.

Principal Component Analysis:
Principal component analysis [7][17] is typically used to reduce the dimensionality and/or to extract new uncorrelated features from the original data. Principal component analysis involves an Eigen analysis on a covariance matrix. If the input data is represented as a matrix **X** of 'n' rows and 'm' columns:

$$X = \begin{bmatrix} x_{11} & x_{12} & .... & x_{1m} \\ x_{21} & x_{22} & .... & x_{2m} \\ . & . & . & . \\ . & . & . & . \\ x_{n1} & x_{n2} & ... & x_{nm} \end{bmatrix}$$

Where $n$ = total number of classes and $m$ = total measures, when finding most effective measure and $n$ = total number of measures and $m$ = total number of classes, when finding the less responsive class. Then, the sample mean $\mu_i$ is computed for each column when finding most effective measure, where

$$\mu_i = \frac{1}{n} \sum_{i=1}^{n} x_{ij} \quad \text{for } j = 1, 2...m.$$

Then X can be centered to form $X^*$

$$X^* = \begin{bmatrix} x_{11} - \mu_1 & x_{12} - \mu_2 & ... & x_{1m} - \mu_m \\ x_{21} - \mu_1 & x_{22} - \mu_2 & ... & x_{2m} - \mu_m \\ . & . & ..... & . \\ . & . & ..... & . \\ x_{n1} - \mu_1 & x_{22} - \mu_2 & .... & x_{nm} - \mu_m \end{bmatrix}$$

Then covariance matrix $R = (1/n)[X^*]^T X^*$ is computed. An Eigen analysis on the covariance matrix R yields a set of positive Eigen values $\{\lambda_1, \lambda_2, ........., \lambda_m\}$. If the Eigen values are sorted in descending order (i.e., $\lambda_1 > \lambda_2 > ..... > \lambda_m$), their corresponding Eigen vectors, $\{v_1, v_2, ...., v_m\}$, are the principal components. The first principal component retains the most variance, if the feature vectors are projected onto the first principal component, more variance will be retained than if the vectors are projected onto any other principal component. The second component retains the next highest residual variance, and so on. A smaller Eigen value contributes much less weight to the total variance. In many cases, the first few components can retain nearly all of the variance. If the 'd' most significant principal components are selected for projection of the data, then the variance (V) retained by this approximation is [8]:

$$V = \frac{\sum_{i=1}^{d} \lambda_i}{\sum_{i=1}^{m} \lambda_i}$$

V is also called the degree of accuracy for the approximation.

## 8. Results and Discussions
### 8.1 Finding an Effective metric for Coupling

For the set of $x_{ij}$ are calculated as defined above for all 3 softwares under observation. Here Software-1 has 20 classes, Software-2 has 11 classes and Software-3 has 13 classes. The Principal Component Analysis produces the principal components for coupling metrics (Shown in Table 1, Table 3 and Table 5).
In case of Software-1 (see Table 1), the projection of the objects into the first principal component retains 76.85% of the total variance and projection of the first two principal components retains 94.99% to the total variance as in Table 2. Observation of the first Eigen vectors reveals that in case of Category1 coupling metrics for Software-1, TNUCC is the most significantly weighted measure.
In case of Software-2 (see Table 3), the projection of the objects into the first principal component retains 85.75% of the total variance and projection of the first two principal components retains 97.76% to the total variance as shown in Table 4. Observation of the first Eigen vectors reveals that in case of Category1 coupling metrics for Software-2, "Class Coupling" is the most significantly weighted measure.
   Table 5 shows the principal components of Category1 coupling metrics of Software-3, the projection of the objects into the first principal component retains 84.46% of the total variance and projection of the first two principal components retains 96.87% to the total variance as shown in Table 6. Observation of the first Eigen vectors reveals that in case of Category1 coupling metrics for Software-3, "Class Coupling" is the most significantly weighted measure.
   The experimental result shows that projecting the object class feature vectors onto the first two principal components retained up to a considerable range of the total variance, hence two components are sufficient to represent the entire dataset with less error. But the most significant component for all these software is not same. Hence there is no uniform coupling measure which remains the most significant component for all software.

**Table 1:** The Principal Components and their Eigen Values for Coupling Metrics of Software 1.

| Principal Component No. | Component Vector | Eigen Value |
|---|---|---|
| 1 | (-0.0388 , 0.0505, 0.0289, **0.7147**, 0.6960, 0.0000) | 2.8132 |
| 2 | (-0.3550, 0.7244, 0.1141, -0.0371, -0.0390, -0.5774) | 0.6641 |
| 3 | (-0.0399 , -0.0228 , 0.0026 , 0.6966 , -0.7160 , -0.0000) | 0.1771 |
| 4 | (-0.4562 , -0.6741 , -0.0521 , -0.0105 , 0.0365 , -0.5774) | 0.0043 |
| 5 | (-0.8112 , 0.0503 , 0.0620 , -0.0476 , -0.0025 , 0.5774) | 0.0018 |
| 6 | (-0.0690 , 0.1236 , -0.9897 , 0.0160 , 0.0118 , 0.0000) | -0.0000 |

**Table 2:** The retained variances of principal components for Coupling Metrics of Software 1.

| Number of Component | % Variance Retained |
|---|---|
| 1 | 76.85% |
| 2 | 94.99% |
| 3 | 99.83% |
| 4 | 99.95% |

Despite of the very interesting research work and studies on coupling measures, there is still a little understanding of the motivation and empirical hypotheses behind many of the measures. It is reported that relating the measures is a difficult task in most of the cases and especially to conclude for which applications they can be used. Analysis shows that there are a lot of inconsistencies in different measures of coupling for object-oriented system. The variations of existing measures reveal that they cannot be represented in a unified framework accepted by all. Therefore, a conclusion can be drawn that the efforts for improving the understanding of object-oriented coupling and for creating a unique framework with empirical validation are useful and necessary.

**Table 3:** The Principal Components and their Eigen Values for Coupling Metrics of Software 2.

| Principal Component No. | Component Vector | Eigen Value |
|---|---|---|
| 1 | (0.0795 , 0.0648 , -0.0225 , -0.5955 , **0.7344** , 0.3084) | 596.6112 |
| 2 | (0.5364 , 0.5395 , -0.0423, -0.4623, -0.3742, -0.2564) | 83.6107 |
| 3 | (0.0373 , -0.0463 , -0.2535 , -0.0989 , -0.4453, 0.8510) | 11.0265 |
| 4 | (0.2428 , -0.8120 , -0.2278 , -0.3757 , -0.1610, -0.2506) | 3.9422 |
| 5 | (0.7729 , -0.1612 , 0.3617 , 0.4093 , 0.1854 , 0.2097) | 0.2327 |
| 6 | (0.2196 , 0.1317 , -0.8665 , 0.3366 , 0.2492 , -0.0910) | 0.3619 |

**Table 4:** The retained variances of principal components for Coupling Metrics of Software 2.

| Number of Component | % Variance Retained |
|---|---|
| 1 | 85.75% |
| 2 | 97.76% |
| 3 | 99.35% |
| 4 | 99.91% |

**Table 5:** The Principal Components and their Eigen Values for Coupling Metrics of Software 3.

| Principal Component No. | Component Vector | Eigen Value |
|---|---|---|
| 1 | (-0.1548 , -0.0351 , -0.2210 , -0.5965 , **0.6043**, -0.4529) | 69.2342 |
| 2 | (-0.3079 , -0.1109 , -0.6325 , -0.3580 , -0.2624 , 0.5438) | 10.1730 |
| 3 | (-0.3139 , 0.0804 , 0.2420 , -0.3704 , -0.6980, -0.4605) | 2.4804 |
| 4 | (-0.6161 , 0.2338 , 0.5561 , -0.0966 , 0.2745, 0.4145) | 0.0583 |
| 5 | (-0.6115 , 0.0560 , -0.3701 , 0.6063 , 0.0554, -0.3394) | 0.0224 |
| 6 | (0.1707 , 0.9603 , -0.2152 , -0.0440 , -0.0199, 0.0037) | 0.0081 |

**Table 6:** The retained variances of principal components for Coupling Metrics of Software 3.

| Number of Component | % Variance Retained |
|---|---|
| 1 | 84.46% |
| 2 | 96.87% |
| 3 | 99.89% |

## 8.2 Finding an Effective class to add a module for extending reusability

Here effective class means the class that has low coupling with respect to others and the class in which if one new module is added then the modification needed will be minimum to achieve the reusability. From the graphical analysis the designer can take the decision of where to add the new module of the existing software in an efficient manner by keeping the development cost as minimum as possible. This leads to a better reusability of the existing software. Here in this we show one example of software 1 (See fig…) where one new module is added to make the existing software reusable. From the graphical analysis of Software 1 the decision that can be taken is: a new module can be added beside the classes which are not highly coupled such as: InformationDialog, CustomButton, ScrollBar, ImagePanel. Also we present a principal component analysis to find a module to interact with for reusing the module.

Table 13 shows the first 3 principal component of software 1. Here first component retains 70.30% variance and first two component retains 95.15% of the total variance. In Table 13 the lower coupling contributing modules are shown in bold face. From the 3 principal components we can see that among the negative impact values only module 9(whose name is InformationDialog and shown as underline) has negative impact to all the principal components. Hence we can take decision if one module is added interacting with only module 9 and then the purpose of the new software serves then it is a good decision to implement it. If the purpose does not serves then we should select a module having low coupling. Also we can see from design pattern of software 1 that the class InformationDialog interacts only with 1 class namely ChatClient. We have implemented our approach for software adding a new class StatusArea and we found that the system is working well serving the new purpose. Figure A shows the design pattern after adding the new module StatusArea.

Table 13: Principal components for finding most Affecting Class (Software 1)

| Principal comp. # | Eigen Vector | Eigen Value |
|---|---|---|
| 1. | (**-0.2190** **-0.0989** 0.5251 0.5679 0.2232 **-0.0705** 0.1037 0.0043 **-0.0430** 0.0168 **-0.0093** 0.0019, 0.0167 0.1911 **-0.1806** 0.3981 **-0.0881** 0.1068 0.0081 0.1707) | 0.9713 |
| 2. | ( 0.9157 **-0.0559** 0.1064 0.1236 0.0472 0.0021 0.0214 0.0033 **-0.0089** 0.0517 0.0505 0.0059, **-0.0289** **-0.1707** **-0.1957** 0.1786 0.0640 **-0.1361** **-0.0267** 0.0150 ) | 0.2244 |
| 3. | (0.0874 0.5140 0.5518 **-0.3007** **-0.1698** 0.0262 0.0652 **-0.0109** **-0.0120** 0.0457 **-0.0055** **-0.1150**, 0.2353 0.1326 0.0021 **-0.2258** **-0.0821** 0.0315 **-0.3285** 0.2104) | 0.0600 |

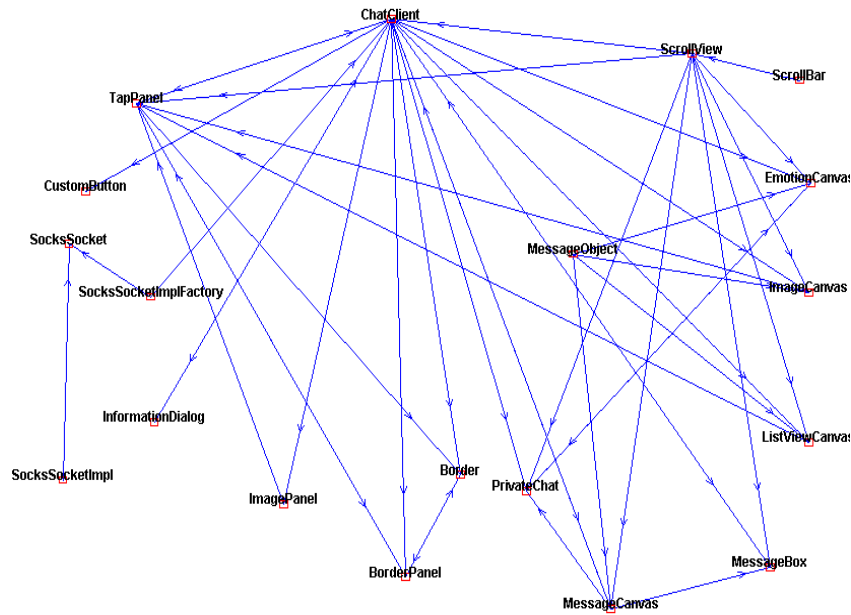

Fig 4 : Graphical Representation of the relationship among the User Defined Classes of software 1(after adding a new module)

## 7. Conclusion

Identification of design patterns from source code is one of the most promising methods for improving software maintainability and reusing design experience. It is hard or even impossible to understand poorly documented legacy systems. Nevertheless, developers try to understand unknown object oriented systems by analyzing the source code to recover the architecture of the system, which is a hard task since the dependencies between the classes cannot be recovered well enough. However when a software of Object Oriented System undergoes the development process, the designer should be concern about the development cost that can be measured with respect to some quality metric of software such as Coupling. In this paper, an approach of detecting design patterns from Java source code is presented and the approach continues with the analysis of source codes of softwares for selecting the most effective component and the highly coupled class. This helps the designer to take the decision that how and where a new module can be added in an efficient manner by keeping the coupling value as minimum as possible and by ensuring the reduction of development cost. We believe, the knowledge about design patterns using the design tool can help developers to understand the underlying architecture faster.

**Authors**

Dr. K. M. Azharul Hasan received his B.Sc. (Engg.) from Khulna University, Bangladesh in 1999 and M. E. from Asian Institute of Technology (AIT), Thailand in 2002 both in Computer Science. He received his Ph.D. from the Graduate School of Engineering, University of Fukui, Japan in 2006. His research interest lies in the areas of databases and software engineering, and his main research interests include Parallel and distributed databases, Parallel algorithms, Information retrieval, Data warehousing, MOLAP, Multidimensional databases, OOAD, Software metric and Software maintenance. He is with the Department of Computer Science and Engineering Khulna University of Engineering and Technology (KUET), Bangladesh since 2001.

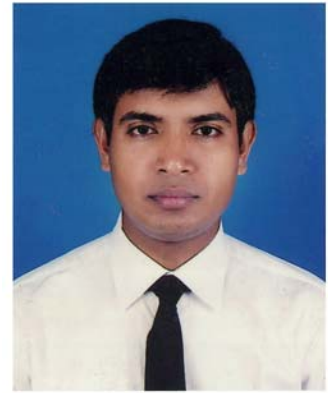

Mohammad Shabbir Hasan received his B.Sc. (Engg.) in Computer Science and Engineering from Khulna University of Engineering and Technology (KUET), Bangladesh in 2008. His research interest includes different areas of Software Engineering like Software Metric, Requirement Gathering, Software Security and Software Maintenance. Currently he is serving as a Lecturer of Department of Computer Science and Engineering in Institute of Science, Trade and Technology (ISTT), Bangladesh.

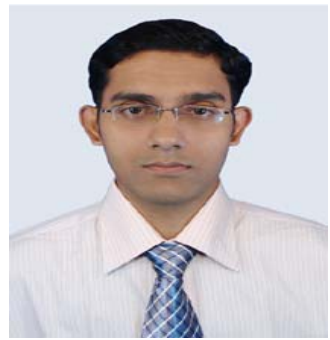